\documentclass[grl]{agutex2015}

 \usepackage{lineno}
\usepackage[dvips]{graphicx}
\setkeys{Gin}{draft=false}  

\authorrunninghead{PALLE ET AL.}

\titlerunninghead{EARTHSHINE ALBEDO 1998-2014}

\begin{document}

%
%

\title{Earth's albedo variations 1998-2014 as measured from ground-based earthshine observations}
%
%

%
%



 \authors{E. Palle\altaffilmark{1,2}, P. R. Goode\altaffilmark{3}, P. Pilar Monta\~n\'es-Rodr\'iguez\altaffilmark{1,2}, A. Shumko\altaffilmark{3}, B. Gonzalez-Merino\altaffilmark{1,2},   C. Martinez Lombilla\altaffilmark{1,2}, F. Jimenez-Ibarra\altaffilmark{1,2},  S. Shumko\altaffilmark{3}, E. Sanroma\altaffilmark{1,2}, A. Hulist\altaffilmark{1,2}, P. Miles-Paez\altaffilmark{1,2}, F. Murgas\altaffilmark{1,2}, G. Nowak\altaffilmark{1,2}, S. E. Koonin\altaffilmark{4 } }


\altaffiltext{1}{Instituto de Astrof\'isica de Canarias (IAC), V\'ia L\'actea s/n 38200, La Laguna, Spain}    
\altaffiltext{2}{Departamento de Astrof\'isica, Universidad de La Laguna, Spain}
\altaffiltext{3}{Big Bear Solar Observatory, New Jersey Institute of Technology,  Big Bear City, CA, 92314, USA}
\altaffiltext{4}{Center for Urban Science \& Progress, New York University, New York City, NY, 11201, USA}





%
%



\noindent \textbf{Key points.}
\begin{itemize}
\item Albedo
\item Earthshine
\item Climate
\item Clouds
\item Aerosol
\item Radiation
\end{itemize}


%
%


\begin{abstract}

The Earth's albedo is a fundamental climate parameter for understanding the radiation budget of the atmosphere. It has been traditionally measured from space platforms, but also from the ground for sixteen years from Big Bear Solar Observatory by observing the Moon. The photometric ratio of the dark (earthshine) to the bright (moonshine) sides of the Moon is used to determine nightly anomalies in the terrestrial albedo, with the aim is of quantifying sustained monthly, annual and/or decadal changes.  We find two modest decadal scale cycles in the albedo, but with no significant net change over the sixteen years of accumulated data. Within the evolution of the two cycles, we find periods of sustained annual increases, followed by comparable sustained decreases in albedo. The evolution of the earthshine albedo is in remarkable agreement with that from the CERES instruments, although each method measures different slices of the Earth's Bond albedo.
\end{abstract}

%
%

%

\begin{article}

\section{Introduction}
The Earth's albedo (or reflectance) is defined as the fraction of solar radiation that is reflected back to space through the top of the atmosphere (TOA).  The global albedo value is $\sim$0.3 in the visible range, and its evolution is controlled by changes in the type and amount of clouds (\citep{2003GeoRL..30.1019C}; \citep{2001JGR...10628371R}), the ice/snow cover and by any changes in continental surface reflectance (\citep{1994JGR....9920757R}). Thus, it is a fundamental regulator of  the energy budget of the planet \citet{2015RvGeo..53..141S}.

There is no long-term historical record of the Earth's albedo, but in the past two decades there have been a number of experiments designed to carefully measure this parameter and its spatial and temporal variability across the globe. From space, several instruments including ERBE (\citep{1989BAMS...70.1254B}) or ScaRab (\citep{1998BAMS...79..765K}), began to provide detailed albedo maps and various global measurements. With the launch of the CERES (\citep{1996BAMS...77..853W}) instruments in 2000, a more systematic and precise monitoring has become possible. CERES measures the  near retroflection of each small area on the Earth twice a day, and interpolates to obtain the visible light Bond albedo (light reflected in all directions) with bi-directional reflectance models.  Beyond this, all space-based determinations rely on absolute measurements, which in the long run might be prone to instrumental drift \citet{2012JGRD..117.5114J}.

Measuring and calibrating earthshine is a complementary method of measuring the Earth's globally-averaged reflectance.  This method does not suffer from long-term calibration issues.  Earthshine was first explained by Leonardo DaVinci (c. 1510), as sunlight reflected from the daytime Earth to the dark side of the Moon and back to the nighttime Earth -- providing an instantaneous, large-scale measure of the Earth's reflectance.  Our observations of the earthshine are made from Big Bear Solar Observatory (BBSO) in California using modern photometric techniques, but built on the century-old pioneering ideas of \citet{1928LAstr..42..572D}.   

Here we report the monthly, annual and decadal variations of this critical climate parameter from terrestrial measurements of the Earth's albedo covering from December 1998 through December 2014 using observations of the earthshine.  This is the first update since 2007.  
 
\section{The Earthshine}
 For about half the nights of each month, we can precisely measure the earthshine of the dark side of the Moon. Half of those nights have lunar phase values progressing over a few nights  from $-150^{\circ}$ to $-50^{\circ}$, as the Moon waxes from new to somewhat more than half full ( with observing nights beginning shortly after sunset have a negative lunar phase angle, and the earthshine-contributing area is located west of BBSO).  For the other half of the monthly observing nights (lunar phases progressing from $50^{\circ}$ to $150^{\circ}$), the Moon wanes  ( with nights ending shortly before sunrise, and the earthshine-contributing area is located east of BBSO).  Each night  yields a single apparent albedo covering $\sim$40\% of the Earth (Pall\'e et al. 2009). Further,  during each night of a waxing (waning) Moon,  we measured an apparent albedo for nearly the same terrestrial area west (east) of BBSO.  Thus, the earthshine observations from a given site provide apparent albedos covering the same $\sim$80\% of the Earth.  

The nightly apparent albedos can be integrated over a month to obtain a measure of the Earth's visible Bond albedo. Our method of collecting and reducing visible light earthshine data is detailed in two lengthy papers (\citet{2003JGRD..108.4709Q}    and \citet{2003JGRD..108.4710P}) and the results of  the data analyses were given in several papers covering observations up to 2007 (\citep{2004Sci...304.1299P}, \citep{2005GeoRL..3211803P}, \citep{2005GeoRL..3221702P}, \citep{2009JGRD..114.0D03P}).

In each night's determination of the apparent albedo, we measure the relative air mass, weighted intensities of the fiducial regions (in the vicinity of Crisium and  Grimaldi on opposing  edges of the Moon) in the dark side (earthshine) and bright side (moonshine) of the Moon.  Their ratio yields a calibrated apparent albedo , which is what we use for the earthshine albedo through the paper.  In detail, and following Qiu et al. (2003) and Palle et al. (2003), the apparent albedo, $p^*$ is 

\begin{equation}
p^*(\beta) = \frac{3}{2f_L(\beta)}\frac{p_bf_b(\theta)}{p_af_a(\theta_0)}
\frac{I_a/T_a}{I_b/T_b}\frac{R^2_{em}}{R^2_{e}}\frac{R^2_{es}}{R^2_{ms}},
\label{eq1}
\end{equation}

where $p^*$ is the albedo of a Lambert (perfectly diffusing) sphere that would have the same
instantaneous reflectivity as the true Earth at the same phase
angle, and where $\beta$ is the Earth's phase angle (the angle between the sunlight that is
incident on the Earth and then reflected to a fiducial region on the dark side of the Moon), $\theta$ is the lunar 
phase angle (the Moon's selenographic phase angle, which is the angle between the direction to the Sun from a  lunar fiducial region and the nighttime observer at BBSO), $\theta_{\circ}$ is 
the small angle between the path of the Earth light striking the lunar fiducial region and the path of the  nearly retroflected earthshine  to the BBSO observer,   ${I_a/T_a\over I_b/T_b}$ is the ratio of the earthshine
intensity ($I$) to the moonshine intensity in two opposing fiducial
patches, after each is corrected by the observed atmospheric transmission ($T$) or airmass, and ${p_b\over p_a}$ is 
the known ratio between the geometrical reflectivity of the two 
opposing fiducial patches. $R_{em}$, $R_{es}$, $R_{ms}$ and $R_{e}$
refer to the Earth-Moon distance, the Earth-Sun distance, the Moon-Sun
distance and the Earth's radius, respectively.  In terms of the Moon's phase angle, the Earth's phase angle, $\beta$, is given by $\beta\approx\pi-\theta$.  The waxing Moon rises 
from $\theta = -\pi$ to 0 and the waning Moon declines from $\theta = 0$ to $\pi$. 
The Moon's phase function for the bright side,
f$_b(\theta)$, is used in the formula to account for the
geometrical dependence of the reflectivity of the Moon, while
$f_a(\theta_0)$ accounts for the fact that the earthshine is not
exactly retroflected from the Moon (${\theta_0}^<_{\sim}
1^{\circ}$).  $f_a(\theta_0)$ is measured using total eclipses of the Moon, and is sensitive to the ``opposition effect'' or enhanced backscatter from the lunar surface, but we are interested in the trends in the apparent albedo, which are insensitive to the opposition effect.  

The Earth's Lambert phase function is given by

\begin{equation}
f_L(\beta) = \frac{(\pi - |\beta| cos\beta ) + sin\beta}{\pi} ,
\label{eq2}
\end{equation}
where
the Earth's phase function is observed to be very roughly
Lambertian for $\mid\beta\mid\le {2\pi\over 3}$ (or $\mid\theta\mid\ge {2\pi\over 3}$).  
In terms of Equations (1) and (2), the  Bond albedo is given by
\begin{equation}
 A={2\over 3}\int_{-\pi}^{\pi} d\theta p^*(\theta) f_L(\theta)\sin \theta 
\label{eq3}
\end{equation}
 Again, for a detailed discussion of these equations and the location of our fiducial patches on the Moon, see \citet{2003JGRD..108.4709Q} and \citet{2003JGRD..108.4710P}.  Note that the lunar phase function was first derived by {\it Hapke} [1963]

\subsection{Earthshine database}
Our primary purpose is to probe the evolution of the  Earth's albedo over time.
Here we present and interpret monthly, annual and decadal anomalies over sixteen years of BBSO earthshine data -- the most recent seven years of which are first published here and serve to clarify trends.  We note that there is a single gap in the entire data string (between November 2005 and August 2006), which was  due to a suspension of observations during dome construction at BBSO for a new solar telescope.

Between December 1998 and January 2007, earthshine observations were made from a telescope under the large dome in Big Bear.  In 2005, a new small dome for earthshine was constructed near the large dome and a new, small, automated earthshine telescope was constructed, installed and tested.  From September 2006 through January 2007, both telescopes were run simultaneously for calibration purposes.  The clearest ten nights covered a wide range of absolute lunar phase angles (70-140 degrees) and showed an excellent agreement where the difference between the two apparent albedos for each night is  nearly the same, small constant offset (independent of lunar phase).  Then an identical automated earthshine telescope was built so that one of the twins would reside at our earthshine station in Tenerife.  The twin telescopes were run side-by-side under the small dome in Big Bear  between July and September 2007 with the fifteen clearest nights covering lunar phases between 80 and 150 degrees.  Again, the difference between the two sets of data for each night could be accounted for by a single, but now appreciably smaller offset.  

Modeling of the telescopes led us to conclude that the primary reason that the data from the original earthshine telescope is the most distinct of the three telescopes is largely due to  the cameras they employed.  The primary difference being the different spectral coverage of the original earthshine telescope camera compared to the latter models (which use identical cameras).  The significant difference in optical design plays a secondary role.  In the Fall of 2007, the first automated telescope was shipped to the Canaries and put into operation under an identical small dome to that in Big Bear.  The inter-calibrations put all the earthshine data on the same footing, so we can determine trends in the apparent albedo, but one should bear in mind that there is a small overall uncertainty in the Bond albedo because of incomplete knowledge of the opposition effect.  This gap will be reduced, {\it post facto}, as more total eclipse data are acquired.  We emphasize that this uncertainty is not present in the differences with respect to the mean and, thus, in any of the trends that are our topic here.   The current knowledge of the opposition effect was discussed in detail in Section 4.4 of Qiu et al. [2003].

From 1998-2006, the earthshine camera was tested annually against a known source for drift and none was found.  In 2006 and 2007, the new cameras were calibrated against each other and the original, and now retired earthshine camera.  The newer cameras have external shutters instead of an extra moonshine filter because the old camera could not take sufficiently short exposures without an additional filter to block most of the moonshine.   Our concern was the possible drift of moonshine filter that had to be added to block most of the signal of the bright side of the Moon to compensate for the inability of the old camera to take very short exposures.

\subsection{Data quality analysis}
In previous works, we typically calculated the 3$\sigma$ bounds for the whole earthshine dataset plotted against absolute value of the lunar phase, and we then sigma-clipped the data to remove outlier values lying more than 3$\sigma$ away from the fit to the data.  The clipping was done by fitting all the data and then eliminating all data more than 3$\sigma$ away from the fit.  Then the fit was re-done and the procedure repeated until the iterations converged, which typically took 2-3 iterations to complete the clipping. If we apply the same methodology to our entire (1998-2014) dataset, more than 90\% of the nights survive the culling (1108 out of 1198).  The nights that do not survive the culling are typically from short observing nights and/or are characterized by weather variations in Big Bear that result in sparse temporal coverage. This methodology leads to albedo anomalies in the early years that are exactly equal to the ones reported in previous papers, but now the analysis is extended up through 2014.

Nevertheless, we noted in \citet{2009JGRD..114.0D03P} that because of an incorrect treatment of outlier data points in \citet{2004Sci...304.1299P}, we reported an overly large trend during 1998-2003.  Our now much longer dataset motivates us to test  more stringent quality control flags to our data. In particular, we have chosen here to test further lowering the acceptance of nights to those whose values do not differ from the mean fitted values by more than 1$\sigma$. If we apply this criterion to our original 1198 measurements, we would be left with the best 678 nights (56\%). This has an apparent, modulating impact on our albedo variations over the long term in our dataset, and leads to a reconciliation of earthshine data with satellite albedo records. We discuss this matter in detail in section 3. 

In Figure~\ref{fig1}, we plot the apparent albedos for the entire sixteen year data set with positive and negative lunar phase angles bundled together and plotted against the absolute value of the phase angle.  The dashed lines in the figure represent  the 1$\sigma$  standard deviation of the mean bounds.  The larger magnitude phase angles correspond to a ``near-crescent'' Moon for which the Earth is nearly Lambertian and the apparent albedo is roughly comparable to the Earth's visible light Bond albedo ($\sim$0.3). 

\section{Long-term albedo trends}

\begin{figure}
\noindent\includegraphics[width=9.3cm]{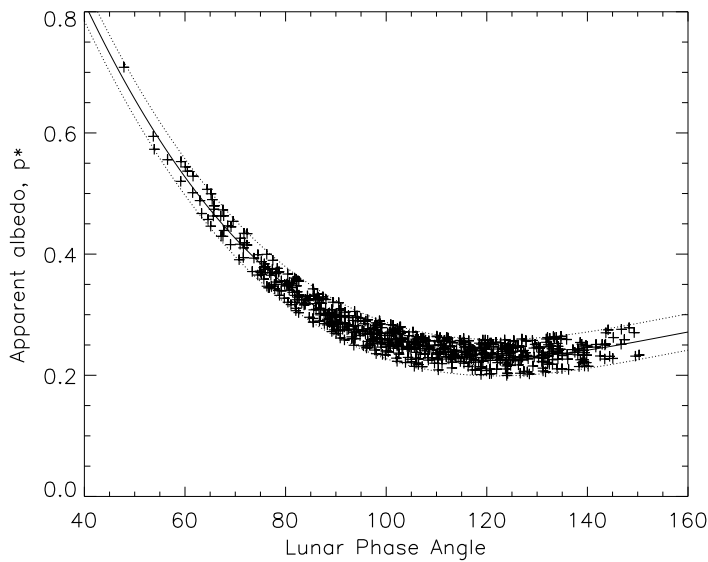}
\caption{Nightly measurements of the apparent albedo from the BBSO earthshine station covering 1998-2014 are plotted as a function of the absolute value of the lunar phase (lunar phase angle = 0 at the full Moon). The fit to all the data is indicated by a thick black line, and the dotted lines enveloping the data are the $1\sigma$ standard deviation from that mean. The fit is a third order polinomial of the form $c_1+c_2 \theta+c_3 \theta^2+c_4 \theta^3$, where $\theta$ is the lunar phase angle and $c_i$ are constants. 
}                                                                                              
\label{fig1}
\end{figure}

In our search for trends in the apparent albedo, we de-seasonalized the entire sixteen year data set.  For this, we separated the apparent albedos of Figure~\ref{fig1} by month and into positive/negative phase angles to constrain the seasonal climatology.  We do not see any significant seasonal variations either for negative or positive lunar phases (not shown). Nevertheless, for consistency, we carry out the de-seasonalization of the data to investigate long-term trends, even though the results reported in this paper are insensitive to whether or not the  de-seasonalization is performed. 

\begin{figure}
\noindent\includegraphics[width=15.0cm]{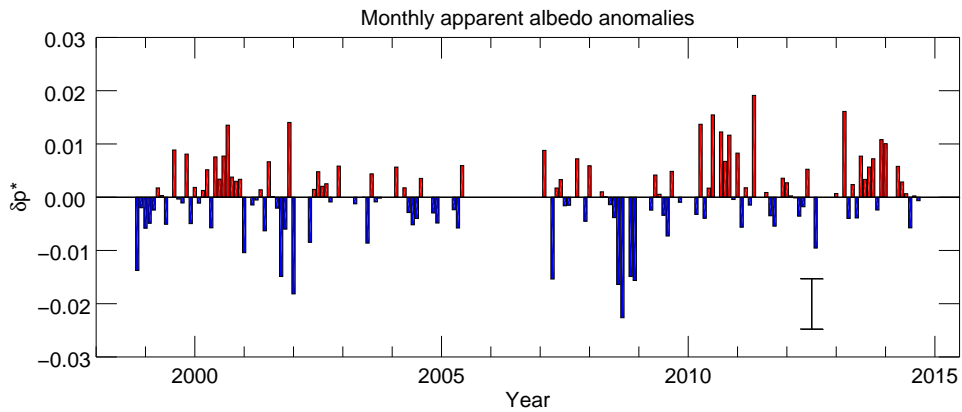}
\caption{Monthly mean apparent albedo anomalies from December 1998 through December 2014. Anomalies were calculated over the mean of the full period, positive anomalies are shown in red and negative in blue. Averaged standard deviation (error) of the monthly data is also indicated in the lower right corner for simplicity. Only months with at least 5 nights of observations are shown. Prior to the calculation of the monthly mean apparent albedo anomalies, the data were  deseasonalized. From November 2005 to August 2006 several months of earthshine data are missing due to the replacement of the dome of the solar telescope, while the new automated telescope  under a separate dome was not yet operational.}                                                                                              
\label{fig3}
\end{figure}

Figure~\ref{fig3} shows the de-seasonalized monthly mean albedo anomalies over the period 1998-2014. Note that we require a minimum of 5 nights of data in a month to include that month in the mean.  Thus, there are a number of months with no value shown in the figure. Over the full sixteen years of our observations, there is no significant long-term trend. Nonetheless, there are a few notable features.  

In particular, Figure~\ref{fig3} roughly indicates two decadal scale oscillations $--$ each of which nets no albedo anomaly.  On shorter time scales, large inter-annual  albedo changes are observable, for example over the 2008-2010 interval there is a strongly increasing trend in albedo.  However, this trend begins to reverse after 2010.  Thus, taken in isolation, over the past several years, 2007-2014, the Earth's albedo, as registered by earthshine, has shown a net trendless interval.

\begin{figure}
\noindent\includegraphics[width=18cm]{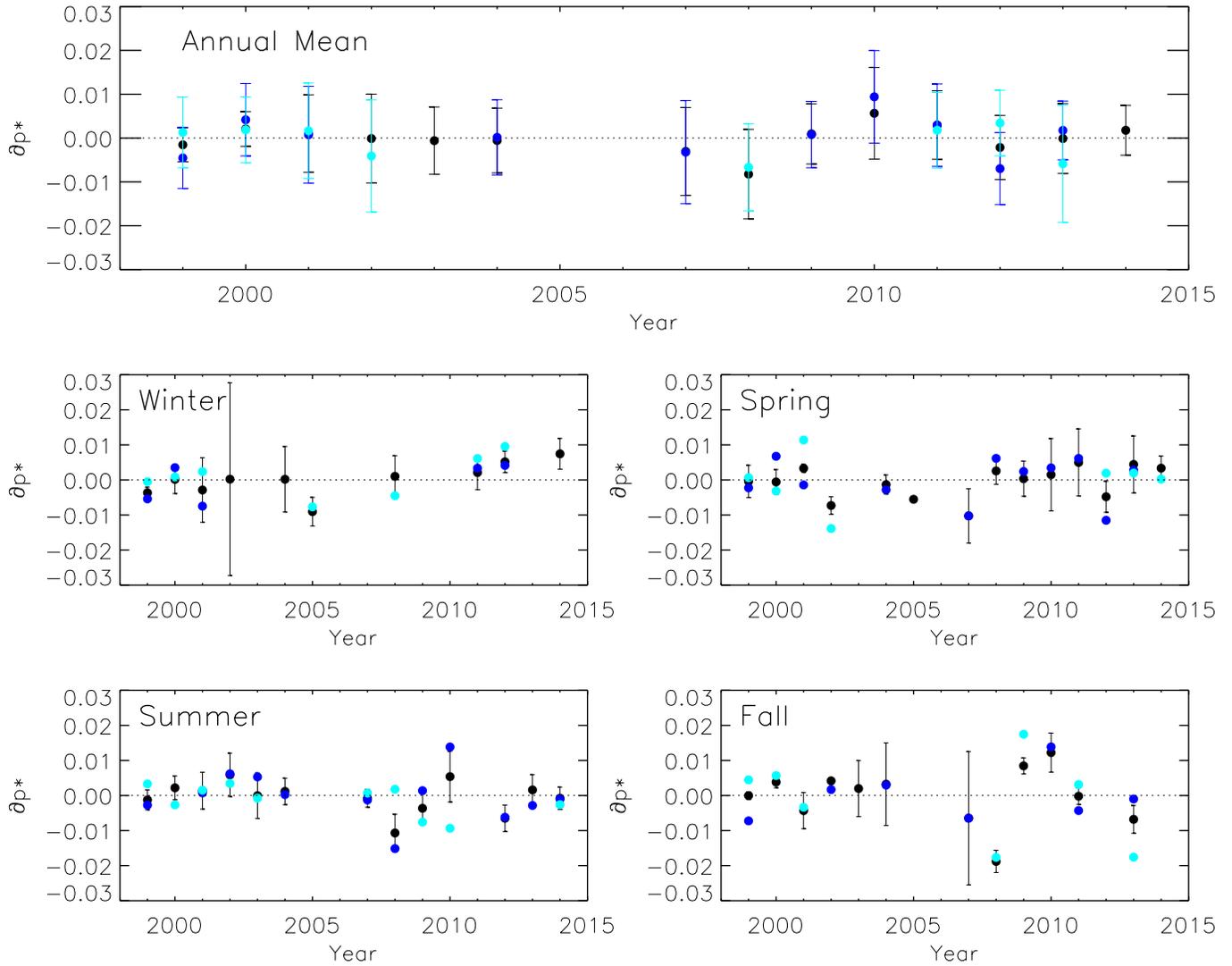}
\caption{Caption: Mean annual and seasonal albedo trends, 1998-2014, from earthshine observations from BBSO.  Black points represent the annual average albedo deviation from the mean.  The blue and cyan points represent annual data from the positive (east-looking) and negative (west-looking) lunar phases, respectively. }                                                                                              
\label{fig4}
\end{figure}

Looking deeper, our data consist of observations of two large, distinct, regions of the Earth, and the combined anomalies in Figure~\ref{fig3} could result from regional changes in either one or both areas with time, which (may) result in a change in the relative contributions of each area to the global average. Further, one might imagine that local weather/climate changes at BBSO could also introduce artificial changes and/or trends in the data via wandering seasonal or sampling favoritisms. To rule out these possibilities, we  separated out  the annual and seasonal means for positive and negative lunar phases. 
The results are plotted in Figure~\ref{fig4}, where we show the sixteen years of albedo anomalies in the upper, large panel and their seasonal decomposition in each of the lower four panels.  For each of the five panels, the geographic decomposition is shown with positive phase values, negative phase values and the combination of both. We are re-assured by the fact that all four plots of the seasonal anomalies are in good agreement among themselves by showing similar patterns. 

The annual means are shown in the larger upper panel.  The long-term anomalies of the two large areas of the earth visible from BBSO (positive and negative lunar phases) vary together, implying that the detected anomalies are global.  

\begin{figure}
\centering
\noindent
\includegraphics[width=9.3cm]{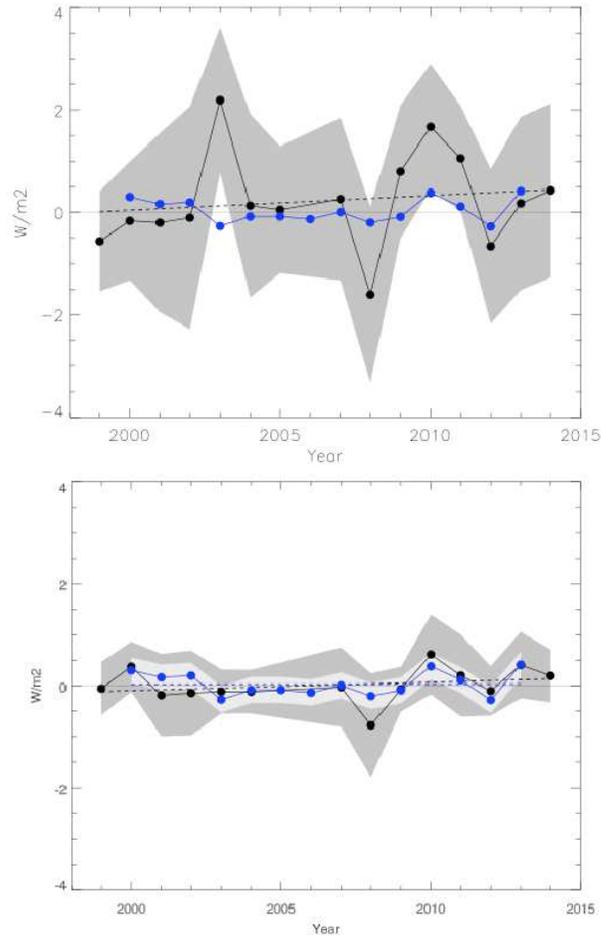}  
\caption{Earthshine annual mean albedo anomalies with 3$\sigma$ (1$\sigma$) clipping in the top (bottom) panel expressed as reflected flux in $W/m^2$.  The bottom panel is a translation of the earthshine annual 
mean anomalies in the top panel of Figure~\ref{fig4}.  The error 
bars are shown as shaded areas and the dashed gray line shows a linear fit to the annual reflected  energy flux anomalies.  
The CERES annual results (2000-2013) are also shown, in blue, with the 
error bars shown as lightly shaded areas in the lower panel only. A linear fit to the CERES data is shown with a blue (lower panel only) dashed line. }
\label{fig5}
\end{figure}

In Figure~\ref{fig5}, we compare trends in Bond albedo anomalies from earthshine data (1998-2014) to CERES data (2000-2013)\footnote{http://ceres.larc.nasa.gov/}.  The ground-based and satellite approaches make different assumptions beyond the data to determine the evolution of the Bond albedo.   For the earthshine, we determine the fractional annual effective albedo anomalies by calculating average fraction, ${\Delta p^*\over p^*}$, where each component of the average is the difference between a measured night in the year of interest and the overall fit to $p^*$ for that phase angle as shown in Figure~\ref{fig1}.  The connection to the Bond albedo anomalies would simply be ${{\Delta p^*}\over {p^*}} = {\Delta A\over A}$, and this would be the connection to the CERES Bond albedo  if earthshine and CERES observations covered the same parts of the Earth (ignoring differences in data collection and analyses, like the clipping done for earthshine). 
Now the changes in albedo are straightforwardly  translated into changes in reflected energy flux at the TOA in $W/m^2$ (see Pall\'e et al. 2004 for details). The top panel, Figure~\ref{fig5}a, shows the earthshine albedo anomalies derived from a 3$\sigma$-clipping criterion, consistent with what was used in previous publications.  The lower panel shows the same set of anomalies, but  derived from a 1$\sigma$-clipping criterion, as shown in Figure~\ref{fig1}, with the result being a much more muted variability after more data are discarded. In both panels, the TOA short-wave reflected flux anomalies from CERES data \citet{2012SGeo...33..359L} are plotted. It is striking how the application of the strict 1$\sigma$-clipping criterion reconciles the measured earthshine albedo anomalies to those derived from CERES space missions. Despite the fact that each annual earthshine mean now contains fewer data points, the size of the error bars  decrease, indicating that a significant part of the eliminated data have a Gaussian (white noise) distribution.   That is, the additional clipping mostly eliminates outliers rather than reducing the number of points contributing to  a real signal.  The choice of 1$\sigma$ for the clipping is somewhat arbitrary, but the result in Figure~\ref{fig5}b is insensitive to modest variations in the choice of bandwidth of $\sigma$-clipping.

It is also noticeable that the clipping affects the amplitude of the measured earthshine anomalies over the whole period 1998-2014. However, the shape of the anomalies is only modified in the earlier period, 1998-2005, before our automated telescope network. In the 2007-2014 period the anomalies persist no matter the choice of  clipping criteria. This can be partly explained by the fact that earthshine observations with the manual telescope usually covered more extreme lunar phases for which accurate measurements were difficult to obtain, either because of scattering from the lunar bright side or by the short duration of the lunar visibility from BBSO. These noisier datapoints tend to deviate more from the mean and thus are more affected by the clipping criteria.   One consequence is the 2003 peak in Figure~\ref{fig5}a is gone from Figure~\ref{fig5}b.  

From Figure~\ref{fig5}b, we emphasize that not only do the earthshine and CERES albedo anomalies agree in magnitude over the common period, but they both show identical structures $--$  both showing a slight decline for the first few years, both the robust dip in 2008, the robust peak in 2010 and turn up at the end of datasets.     The agreement is even more impressive considering the distinctly different components of the Bond albedo to which each measurement is sensitive.  

Two related and striking issues are the good agreement between the earthshine and CERES anomalies in the face of rather large earthshine error bars  and the question of where should the $\sigma$-clipping end.  We find that we can continue to decrease the band width of the $\sigma$-clipping down to about 0.6~$\sigma$ at which point so much of the data ($\approx 70\%$) have been eliminated that further clipping leaves meaningless results (not enough points to make statistically significant monthy or yearly means).  However, even at 0.6~$\sigma$ the error bars are further reduced, while the earthshine results remain largely unchanged from the results for 1~$\sigma$, which implies the shaded earthshine error bars in Figure~\ref{fig5}b are probably overestimated.  

In summary, \citet{Wielicki06052005} reported a strong disagreement between earthshine and CERES albedo anomalies, over the initial, short period of 2000-2003. This disagreement was reduced when updated CERES albedo anomalies became available after their improved recalibration of the on-board sensors \citet{2007JCli...20..575L}. With our new, more strict clipping of the earthshine data, and extending the comparison to the full interval 2000-2013, the differences between CERES and Earthshine data have practically disappeared.

\section{Discussion and Conclusions}

We have reported here a new 16-year long dataset of the Earth's nearly globally-averaged albedo as measured by looking at the earthshine. Monthly and yearly means are presented after we applied a strict criteria for the acceptance of data points.  The Earth's reflectance as measured from earthshine is noticeably variable on monthly, yearly and decadal scales with a net over sixteen years that is essentially nil, but with decadal and inter-annual variabilities.

After CERES recalibrations and a stricter criterion for acceptable earthshine data, both sets of albedo anomalies are in good agreement over the 14 years they have in common.  The agreement is carried into every annual trend, ending a decade long discussion on the nature of the disagreements. The peaks and valleys in the albedo anomalies seems real, but of unclear origin. Continuation of the albedo measurement over a longer period of time, and during large synoptic episodes such as strong El Ni\~no/La Ni\~na events will help to resolve these issues, as well as pinning down the role of large-scale phenomena.

Despite the agreement of the two datasets, the error bars of the earthshine measurements are noticeably larger than those of the CERES data. To further significantly improve the precision of the earthshine data, one would need a global network.  At the moment, two automated stations $--$ one in Big Bear and one in Tenerife $--$ are collecting measurements regularly.  With eight small robotic stations uniformly spaced in longitude, we would be able to obtain global coverage with about 2-3 times the precision we have now.  The cost and effort in robotic telescope have both greatly reduced in recent years.  Such a global network would have overlapping coverage, enabling differential sampling from the stations, whereas now we have non-redundant coverage from BBSO with $\sim$80\% of the Earth between the poles. Another solution would be the use of a simple nanosat to measure continuously the Moonshine/Earthshine ratio from space. Both options provide a way to obtain an absolutely calibrated albedo time series for climate changes at a very reduced cost. 

We also notice that not only are the broadband albedo measurements  relevant to the Earth's radiation budget, but also to its wavelength dependence/variability \citet{2005ApJ...629.1175M}.  On the Tenerife earthshine telescope, we have installed an automated filterwheel for this purpose, which is now beginning operation, and we plan to provide the first long-term times series of color albedo measurements.


%
%
%
%
%
%
%

\begin{acknowledgments}
Monthly and yearly Earthshine measurements are available upon request to the authors. This work is partly financed by the Spanish Ministry of Economics and Competitiveness through projects ESP2013-48391-C4-2-R and ESP2014-57495-C2-1-R.
\end{acknowledgments}

\end{article}

%
%
%
%
%

%


\end{document}